\RequirePackage{lineno}
\documentclass[prb,twocolumn,superscriptaddress,preprintnumbers,amsmath,amssymb,floatfix]{revtex4-2}

\usepackage{graphicx}
\usepackage{color}
\usepackage{lineno}
\usepackage{mathrsfs}

\begin{document}

\title{Plasma echoes in graphene}

\author{Marinko Jablan}
\email{mjablan@phy.hr}
\affiliation{Department of Physics, University of Zagreb, 10000 Zagreb, Croatia}

\date{\today}

\begin{abstract}
Plasma echo is a dramatic manifestation of plasma damping process reversibility. In this paper we calculate temporal and spatial plasma echoes in graphene in the acoustic plasmon regime when echoes dominate over plasmon emission. 
We show an extremely strong spatial echo response and discuss how electron collisions reduce the echo. We also discuss differences between various electron dispersions, and differences between semiclassical and quantum model of echoes. 
\end{abstract}

\maketitle

\section{Introduction}

Spin echo denotes a peculiar response when a spin ensemble is excited by two separated pulses. 
While system response decays after each pulse has ceased, after some time our system spontaneously regroups and shoots out another pulse, indicating a reversible dynamics \cite{Hahn50}. This is a generic behavior that can be understood from the simple example of the Hahn's horse herd. After a cowboy shoots a gun horses will run away, and since each has a different velocity our herd will spread. If after some time another cowboy in front of the herd shoots his gun, horses will turn around and at particular instant regroup at the initial herd size (here we neglect horse collisions and acceleration times). 
Similar effect occurs in electron plasmas \cite{ONeil68} and it is the purpose of this paper to explore the echo effect in the two-dimensional (2D) graphene electron plasma \cite{Jablan09}. 

We describe our plasma by a semiclassical distribution function $f({\bf r},{\bf p},t)$, where ${\bf r}=x\hat{\bf x}+y\hat{\bf y}$ is radius vector and $\bf p$ is electron momentum in the 2D graphene plane. 
In the equilibrium case at temperature $T$ and Fermi energy $\mathscr{E}_F$ we get 
a Fermi-Dirac distribution: $f_0({\bf p})=\frac{4}{h^2}\frac{1}{e^{(\mathscr{E}_{\bf p}-\mathscr{E}_F)/k_BT}+1}$  \cite{LLPK} .
Here $h^2$ is the semiclassical phase space volume per quantum state, and we took into account 2 spin and 2 valley degeneracy in graphene \cite{Wallace47}.
The main point of this paper can be understood from the following simple  analysis. Let us apply two electric field pulses separated by a time $\tau$, both periodic in space: ${\bf E}=A_1\delta(t)\cos k_1x\ \hat{\bf x}+A_2\delta(t-\tau)\cos k_2x\ \hat{\bf x}$. In the linear response regime first pulse will induce a distribution change:
\begin{equation}
\delta f_1\propto A_1 \cos k_1x \cdot \frac{\partial f_0}{\partial p_x}.
\label{linear1}
\end{equation}

Later we will show that the constant of proportionality is just the electron charge, but for now we will only give a rough analysis. For example it is easy to see that in the lowest order we need to have a factor of a type $\partial f_0/\partial p_x$ in the response. Namely if $f_0(p_x)$ was a constant than the electric field would make no effect on our plasma since it would just shuffle electrons $\dot{p}_x=-eE_x$ between states of a featureless distribution. 
After the external pulse has ceased our system continues to evolve as:
\begin{equation}
\delta f_1\propto A_1 \cos k_1(x-v_xt) \cdot \frac{\partial f_0}{\partial p_x},
\label{free1}
\end{equation}
where we assumed that particles in plasma evolve independently which we will justify later. We can now see that at large times $t$ we get fast oscillations in $\delta f_1({\bf p})$ which results in large cancelations in the integral of the induced particle density $\delta n_1=\int d{\bf p} \ \delta f_1 $. In other words we expect to see density decay as time increases. Specifically in 2D case:
\begin{equation}
\delta n_1\propto A_1 \int_0^\infty pdp \frac{df_0}{dp}\int_0^{2\pi}d\theta \cos\theta\cos k_1(x-vt\cos\theta).
\label{density}
\end{equation}

Particularly in graphene we have massles Dirac electrons with linear energy dispersion $\mathscr{E}({\bf p})=v_F |{\bf p}|$ (note a singularity at the Dirac point ${\bf p}=0$), and velocity ${\bf v}({\bf p})=\partial \mathscr{E}/\partial{\bf p}=v_F\hat{\bf p}$ of constant magnitude $v=v_F$ \cite{Wallace47}. We thus obtain the particle density: 
\begin{equation}
{\delta n_1(x,t)\propto A_1\sin k_1x \cdot J_1(k_1v_Ft)},
\label{densityD}
\end{equation}
where $J_\nu$ is a Bessel function of the first kind of the order $\nu$ \cite{AS}, and we only wrote the $(x,t)$ density dependence. Furthermore since $J_\nu(\xi)\sim\sqrt{2/\pi \xi}\cos(\xi-\nu\pi/2-\pi/4)$ for large $\xi$, we see that density decays like $1/\sqrt{t}$ at large times. Note that this result is independent of temperature in the graphene case which is a consequence of the linear electron dispersion i.e. a Dirac point singularity. 

In the parabolic dispersion case $\mathscr{E}({\bf p})=p^2/2m$ at zero temperature we get the same type of behavior as given by Eq. (\ref{densityD}), only now $v_F=\sqrt{2\mathscr{E}_F/m}$. However this behavior can now be traced to the Fermi sea singularity since $df_0/dp\rightarrow\infty$ at the Fermi momentum $p_F=\sqrt{2m\mathscr{E}_F}$.
One obtains very different result in the high temperature case of Boltzman distribution. It is then more convenient to separate the integral (\ref{density}) into Cartesian coordinates $\int dp_xdp_y$, and by performing the saddle-point analysis one finds the exponential decay: $dn_1\sim e^{-k_1^2v_T^2t^2/2}$, where $v_T=\sqrt{k_BT/m}$ \cite{LLPK}.

In all these cases we can see that plasma density decays in time even though we didn't consider any dissipative processes. This plasma decay is thus a perfectly reversible process which can in turn be demonstrated by an echo experiment. To show this we look at the response to the second pulse at a large time $\tau k_1v_F\gg 1$. One again obtains the response of the type shown in Eq. (\ref{linear1}) which decays after the second pulse has ceased. However now there is another response (also linear in the second field amplitude) given by:
\begin{equation}
\begin{split}
\delta f_2&\propto A_2 \cos k_2x \cdot \frac{\partial \delta f_1}{\partial p_x} \\
&\propto A_2\cos k_2x \cdot A_1\sin k_1(x-v_x\tau),
\label{linear2}
\end{split}
\end{equation}
where  we explicitly wrote only the $x$-dependence.
Note that we have used Eq. (\ref{free1}) and the derivative in $\partial \delta f_1/\partial p_x$ has affected only the argument of the cosine since this extracts a large factor $\propto\tau$.
After the second pulse has ceased our system continues to evolve as:
\begin{equation}
\begin{split}
\delta f_2\propto A_1A_2 \sin k_1(x-v_xt)\cos k_2(x-v_x(t-\tau)) \\
=\frac{1}{2}A_1A_2 \sin((k_1+k_2)x-(k_1+k_2)v_xt +k_2v_x\tau) \\
+\frac{1}{2}A_1A_2\sin((k_1-k_2)x-(k_1-k_2)v_xt -k_2v_x\tau).
\label{free2}
\end{split}
\end{equation}
By looking at the argument of the second sine we see that at a particular time: 
\begin{equation}
t=\tau_{echo}=\tau\cdot\frac{k_2}{k_2-k_1},
\label{echot}
\end{equation}
all the $\bf p$-dependence disappears and we remove the cancelations in the integral of $\delta n$ i.e. we get a strong echo response. The same trick clearly doesn't work with the first sine since then echo had to occurred at a time $\tau k_2/(k_2+k_1)<\tau$ i.e. before the second pulse was even launched. The same argument dictates that $k_2>k_1$ since otherwise the echo from Eq. (\ref{echot}) would had to occurred at time $\tau_{echo}<0$ i.e. before even the first pulse was launched.

\section{Plasmon emission}

So far we have neglected particle interactions which are of course very important. For example long range Coulomb interaction can result in collective plasma oscillation (plasmon). Then in addition to the echoes discussed in the previous chapter, our pulsed sources can also excite plasmons. It is the purpose of this chapter to discuss when we can neglect this plasmon emission. 
Let us first neglect particle interactions beyond mean field (we discuss collisions at the end of the paper). Plasma dynamics is then described by the Vlasov equation \cite{LLPK}:
\begin{equation}
0=\frac{df}{dt}=\frac{\partial f}{\partial t}+\dot{\bf r}\cdot\frac{\partial f}{\partial \bf r}+\dot{\bf p}\cdot\frac{\partial f}{\partial \bf p}.
\label{vlasov}
\end{equation}

The force on the electron is: $\dot{\bf p}=-e\bf E$, where ${\bf E}={\bf E}^{ext}+{\bf E}^{ind}$ contains both the externally applied electric field and the induced (screening) field. We focus on the regime where electrons and photons (i.e. polaritons) propagate at velocities $\sim v_F\ll c$ so we neglect relativistic effects and introduce the electrostatic potential via: ${\bf E}=-\nabla\varphi$.

Let the graphene plane sit on a dielectric of permittivity $\varepsilon_1$ and 
width $d$ which in turns sits on the perfectly conducting metal plate. Furthermore we assume a dielectric of permittivity $\varepsilon_2$ filling the space above the graphene. 
External potential will induce charge density $\rho^{ind}=-e\int d{\bf p} \ \delta f$ which will create the potential $\varphi^{ind}$.  
It is straight forward to solve the Maxwell equations for the 2D charge density oscillating at a particular Fourier component: $\rho^{ind}({\bf r})=\rho^{ind}_{\bf k}e^{i{\bf k}\cdot{\bf r}}$. Fourier component of the induced potential is given by:
\begin{equation}
\varphi^{ind}_{\bf k}=\frac{\rho^{ind}_{\bf k}}{2k\varepsilon_0} \frac{2(1-e^{-2kd})}{\varepsilon_1+\varepsilon_2+(\varepsilon_1-\varepsilon_2)e^{-2kd}}
=\frac{\rho^{ind}_{\bf k}}{2k\varepsilon_0\varepsilon_k},
\label{induced}
\end{equation}
where we have introduced the wavenumber dependent permittivity $\varepsilon_k$ for convenience. Particularly if the conducting plate is far ($kd\to\infty$) we simply get the average permittivity: $\varepsilon_k=\bar\varepsilon=(\varepsilon_1+\varepsilon_2)/2$, while if the plate is near ($kd\to 0$) we get: $\varepsilon_k\to\infty$. 

Let us now look at the linear response to the external potential. In that case we can separately treat each Fourier component: $\varphi^{ext}({\bf r},t)=\varphi^{ext}_{{\bf k}\omega}e^{i({\bf k}\cdot{\bf r}-\omega t)}$. From Eq. (\ref{vlasov}) we find Fourier component of $\delta f=f-f_0$:
\begin{equation}
\delta f_{{\bf k}\omega}=-\frac{e\varphi_{{\bf k}\omega} {\bf k}}{{\bf k}\cdot{\bf v}-\omega}\cdot\frac{\partial f_0}{\partial\bf p}.
\end{equation}
We can then calculate induced charge density $\rho^{ind}_{{\bf k}\omega}=-e\int d{\bf p} \ \delta f_{{\bf k}\omega}$, and using Eq. (\ref{induced}) the induced potential $\varphi^{ind}_{{\bf k}\omega}=\varphi_{{\bf k}\omega}-\varphi^{ext}_{{\bf k}\omega}$. We thus obtain the total system response to the external potential:
\begin{equation}
\varphi_{{\bf k}\omega}=\frac{\varphi^{ext}_{{\bf k}\omega}}{\varepsilon(k,\omega)},
\label{screen}
\end{equation}
where we have introduced the dielectric function:
\begin{equation}
\varepsilon(k,\omega)=1-\frac{e^2}{2k\varepsilon_0\varepsilon_k}\int\int \frac{pdp d\theta}{{\bf k}\cdot{\bf v}-\omega}{\bf k}\cdot\frac{\partial f_0}{\partial\bf p}.
\label{eps}
\end{equation}

Note that the dielectric function in any isotropic system depends only on the magnitude of the wavevector $k=|{\bf k}|$, and generally satisfies: $\varepsilon(k,-\omega)=\varepsilon^*(k,\omega)$ \cite{LLPK}. 

Finally let us note that the electric field in the graphene plane: ${\bf E}_{\bf r}=-\nabla_{\bf r}\varphi$, is longitudinal with Fourier components: ${\bf E}_{\bf k}=-i{\bf k}\varphi_{\bf k}$. We will be mostly interested in 1D variations when all wave vectors are along $x$-direction for example. Then it is convenient to introduce the $x$-component of the field: $E_{\bf k}={\bf E}_{\bf k}\cdot\hat{\bf x}$. From Eq. (\ref{screen}) we can then find the total response to the external field:
\begin{equation}
E_{{\bf k}\omega}=\frac{E^{ext}_{{\bf k}\omega}}{\varepsilon(k,\omega)},
\label{screenE}
\end{equation}
or from Eq. (\ref{induced}) we can find the induced field:
\begin{equation}
E_{\bf k}^{ind}=-ik_x\frac{\rho_{\bf k}^{ind}}{2|k_x|\varepsilon_0\varepsilon_{|k_x|}},
\label{indE}
\end{equation}
which is what experimentalist can ultimately measure. 

Particularly in graphene ${\bf k}\cdot{\bf v}=kv_F\cos\theta$ so we can separate integrals over $dp$ and $d\theta$ in (\ref{eps}). Integral over $dp$ can be easily solved by partial integration while the integral over $d\theta$ can be solved using the residuum theorem in the complex plane $\zeta=e^{i\theta}$, so we obtain:
\begin{equation}
\varepsilon(k,\omega)=1+\chi_k\left(1-\frac{1}{\sqrt{1-k^2v_F^2/\omega^2}} \right),
\label{difu}
\end{equation}
where we have introduced:
\begin{equation}
\chi_k=\frac{4\pi e^2 k_BT }{k\varepsilon_0\varepsilon_kh^2v_F^2}\ln\left(1+e^{\mathscr{E}_F/k_BT}\right).
\end{equation}

\begin{figure}
\centerline{
\mbox{\includegraphics[width=0.5\textwidth]{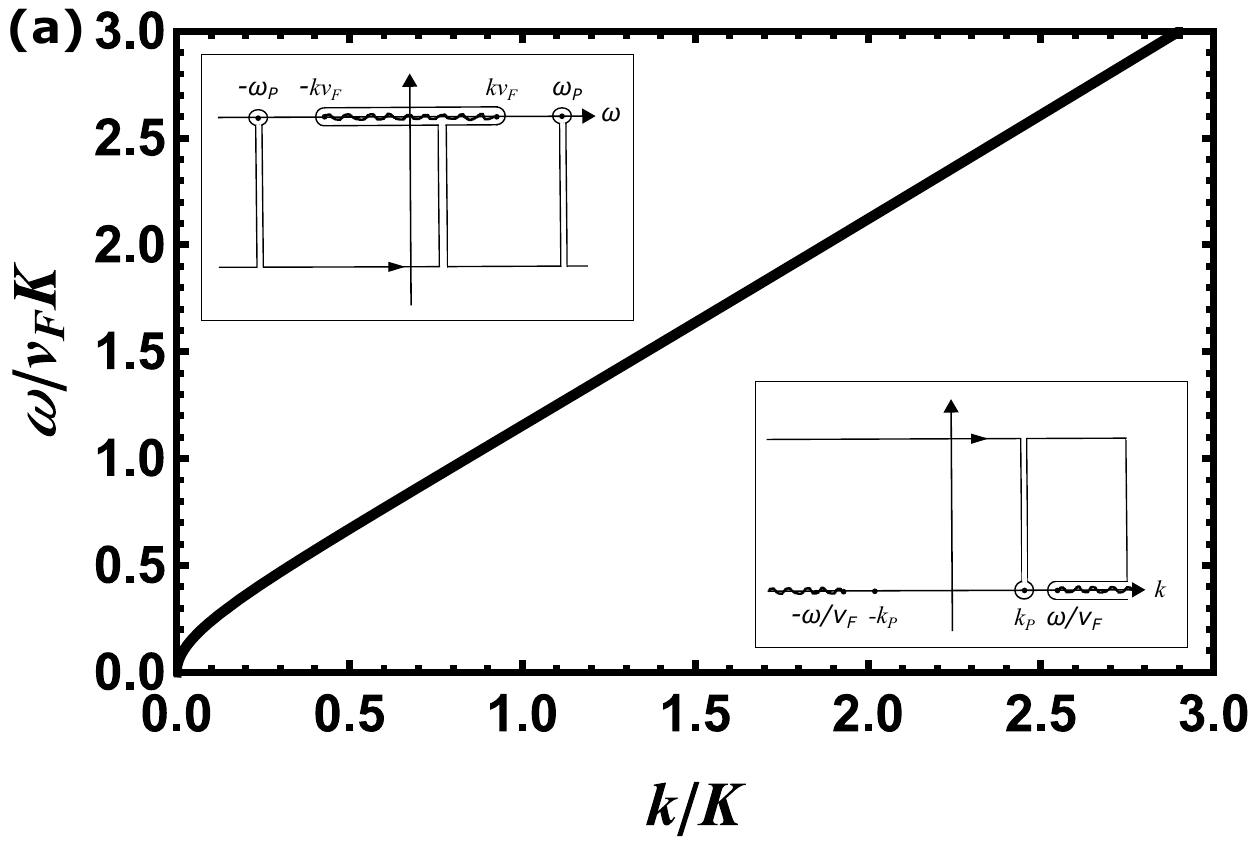}}}
\centerline{
\mbox{\includegraphics[width=0.5\textwidth]{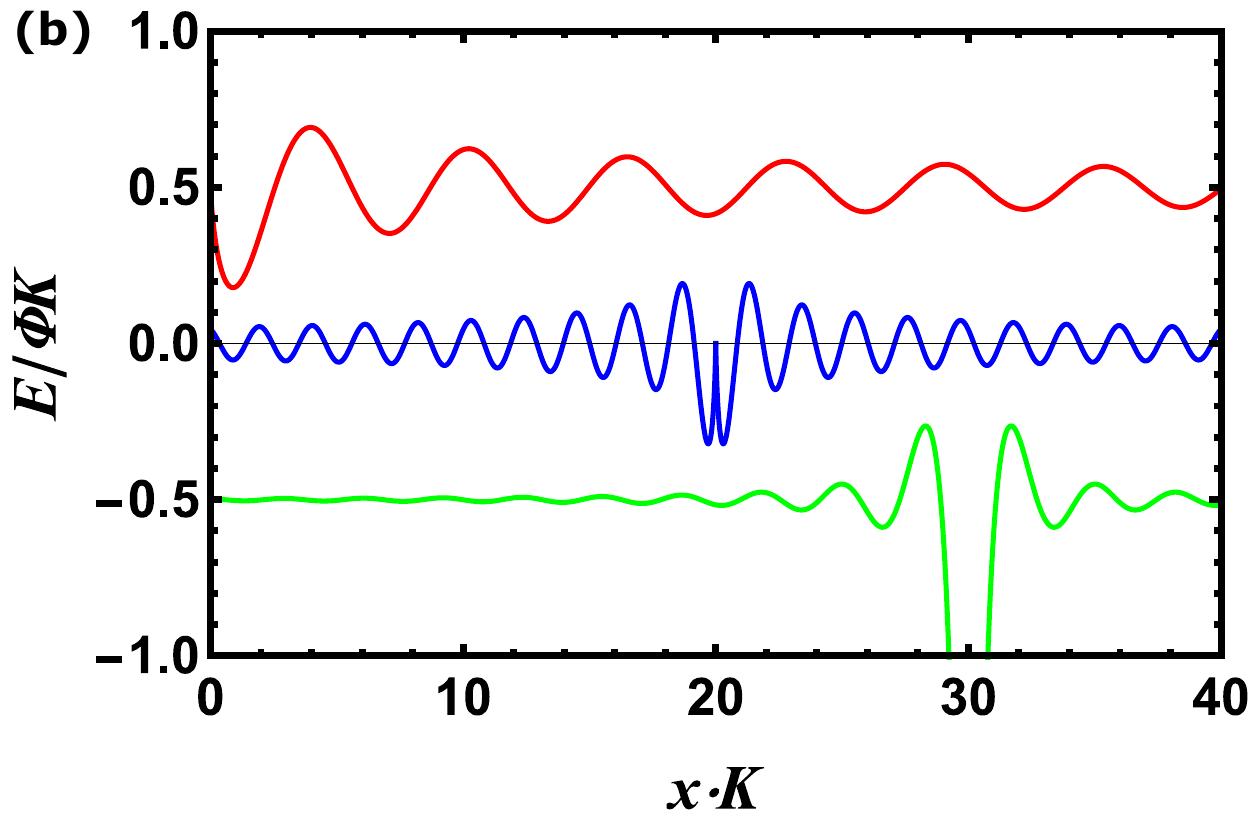}}}
\caption{(a) Plasmon dispersion in graphene. Left (i.e. right) inset shows integration contour in complex $\omega$-plane ($k$-plane) for perturbation localized in time (space) and periodic in space (time).
(b) Plasma response to an electric field perturbation: $E^{ext}(x,t)=-\varPhi\delta(x)\cos\omega_1t-\varPhi\delta(x-l)\cos\omega_2t$, where: $\omega_1=v_FK$, $\omega_2=3v_FK$, $e\varPhi=8\mathscr{E}_F/lK$, and $lK=20$.
We plot each field frequency with different color and shift vertically the red curve by $\varPhi K/2$, and the green curve by $-\varPhi K/2$. Red, blue and green curves present fields oscillating at frequencies $\omega_1$,  $\omega_2$ and $\omega_3=\omega_2-\omega_1$, respectively. Total field is the sum of these three contributions.
Note a divergence of the green echo field at: $l_{echo}=l\omega_2/\omega_3$, signifying a breakdown of a perturbative approach at this point. This divergence is removed by electron collisions (see text for details).}
\label{fig}
\end{figure}

Note that $\varepsilon(k,\omega)$ diverges at the line $\omega/k=v_F$ which is a very specific consequence of the linear Dirac dispersion where all electrons move at the same velocity $v_F$. Further more for $0<\omega/k<v_F$ we have a positive imaginary part of $\varepsilon$ which signifies damping process, while for $\omega/k>v_F$ there is no damping and graphene can support plasmon modes defined by $\varepsilon(k,\omega)=0$. In that case our systems supports free plasma oscillations even in the absence of external field (see Eq. (\ref{screen})). Plasmon dispersion is easily calculated from Eq. (\ref{difu}):
\begin{equation}
\omega_P=kv_F \cdot \frac{1+\chi_k}{\sqrt{1+2\chi_k}},
\end{equation}
which is plotted in Fig. \ref{fig}(a) for the case $\varepsilon_k=\bar\varepsilon$ (absence of metal plate). In that case it is convenient to introduce system scale parameter $K=k\chi_k$ and plot $k$ in these units. 
Generally plasmon reaches almost acoustic dispersion $\omega_P\approx kv_F$ for $\chi_k\ll 1$ which can be obtained with large substrate screening (large $\varepsilon_{1,2}$ or small distance $d$ to the metal plate). However it is easy to check that plasmon dispersion can never cross the singular line $\omega=kv_F$ where $\varepsilon(k,\omega)$ diverges (since $\sqrt{1+2\chi}<\sqrt{1+2\chi+\chi^2}=1+\chi$). Plasmons are then never damped in the semiclassical graphene case, which is very different from the parabolic electrons where plasmons can enter the regime of imaginary $\varepsilon(k,\omega)$ and get damped (the so called Landau damping \cite{LLPK}). However if $\chi_k\ll 1$ plasmon dispersion $\omega_P\approx kv_F$ is very close to the dielectric function singularity, and it was shown that these acoustic plasmons are very weakly excited by a pulse source localized in time \cite{Kukhtaruk15}. We will show that the same is true of the source localized in space. Note that this are bad news for acoustic plasmon nonlinear response which was shown to be extremely large \cite{Jablan20}. In the future papers we will discuss how to efficiently excite acoustic plasmons and fully benefit from these nonlinearities. 

Let us first look at the system response to a pulse localized in time, periodic in space. We apply an electric field in the graphene plane: ${\bf E}^{ext}=E^{ext}\hat{\bf x}$, where:
\begin{equation}
E^{ext}(x,t)=-A\delta(t)\cos kx.
\end{equation}

Note that: $A=-\int E^{ext}(0,t)dt$, which corresponds to the amplitude of the vector potential (in the gauge $\varPhi=0$, $A\neq 0$).
From Eq. (\ref{screenE}) we easily find the total field:
\begin{equation}
E(x,t)=-A\cos kx\int_{-\infty}^{\infty}\frac{d\omega}{2\pi}\frac{e^{-i\omega t}}{\varepsilon(k,\omega)}.
\label{intO}
\end{equation}
This integrand has singularities at the plasmon poles $\omega=\pm\omega_P$ where $\varepsilon(k,\omega)=0$ and at the points $\omega=\pm kv_F$ where $\varepsilon(k,\omega)$ has singularities. To take these points properly into account we adiabatically turn on the external field $\propto e^{\eta t}$, where $\eta\to 0^+$. This amounts to adding a positive imaginary part to the frequencies of the Fourier components: $E^{ext} e^{\eta t}=\int E^{ext}_\omega e^{-i(\omega+i\eta)t}d\omega/2\pi$. This means that now the singularities of $1/\varepsilon(k,\omega+i\eta)$ are pushed into the lower half of the complex $\omega$ plane. In other words, the integral (\ref{intO}) passes slightly above mentioned singularities. 
To evaluate the integral for $t>0$ we close the contour of integration in the lower half of the complex $\omega$ plane (see Fig. \ref{fig}(a)). Result is simply given by the singularities of the function $1/\varepsilon(k,\omega)$, which are given by the zeros of $\varepsilon(k,\omega)$ (i.e. the plasmon poles) and the singularities of the $\varepsilon(k,\omega)$ which we specify by putting a branch cut along the line $\langle-kv_F,kv_F\rangle$. Plasmon poles are easily evaluated by the residuum theorem:
\begin{equation}
E_P(x,t)=2A \cos kx\frac{\sin\omega_Pt}{\partial\varepsilon/\partial\omega |_{\omega_P}}.
\end{equation}
One can immediately see that we can reduce the plasmon emission by using acoustic plasmons where $\partial\varepsilon/\partial\omega |_{\omega_P}$ is large. This is because there is a fast growth of dielectric function from the plasmon pole $\varepsilon(k,\omega_P\approx kv_F)=0$ to the singularity $\varepsilon(k,kv_F)=\infty$ over a short frequency interval. Mathematically if $\chi_k\ll 1$ we get from Eq. (\ref{difu}):
\begin{equation}
\left.\frac{\partial\varepsilon}{\partial\omega}\right\vert_{\omega_P}\approx\frac{1}{\chi_k^2 kv_F},
\end{equation}
so the plasmon field is:
\begin{equation}
\begin{split}
E_P(x,t)&\approx 2\chi_k^2\ kv_F A\cos kx\sin \omega_Pt\\
&\approx 2\chi_k^2\ kv_F A\cos kx\sin kv_Ft.
\label{plasmont}
\end{split}
\end{equation}

Remaining part of the integral (\ref{intO}) is determined by the behavior along the branch cut. We shall show that this is the dominant part when $\chi_k\ll 1$. From Eq. (\ref{difu}):
\begin{equation}
\frac{1}{\varepsilon(k,\omega)}\approx 1-\chi_k\left(1+\frac{i\omega}{\sqrt{k^2v_F^2-\omega^2}} \right),
\label{noscreen}
\end{equation}
which is valid except near the branch points $\omega=\pm kv_F$ which contribute little (since the integral of the type $\int_0^l dx/\sqrt{x}$ is small for small $l$). Note that $\varepsilon$ has an imaginary part in the interval $\omega\in\langle-kv_F,kv_F\rangle$ whose sign can be easily found from the general properties of dielectric function which has to satisfy:  $\Im\varepsilon(\omega>0)\geq 0$, along the physical contour (just above the cut) \cite{LLPK}.
We thus obtain the integral along the cut:
\begin{equation}
\begin{split}
E_D(x,t)&\approx A\cos kx \cdot 2\int_{-kv_F}^{kv_F}\frac{d\omega}{2\pi}\frac{\chi_ki\omega e^{-i\omega t}}{\sqrt{k^2v_F^2-\omega^2}} \\
&=\chi_k\ kv_F A\cos kx J_1(kv_Ft),
\end{split}
\label{diract}
\end{equation}
where the last equation was obtained from the integral representation of the Bessel function \cite{AS}. Since this singularity at $\omega=kv_F$ is a very specific result of the Dirac electron linear dispersion we refer to this mode $E_D$ as the Dirac mode for convenience. We can now clearly see that the plasmon emission is suppressed by a factor $\chi_k\ll 1$ compared to the Dirac mode. However experimentalists should be careful not to confuse these two as both look similar superficially. Indeed both oscillate at similar frequencies $\omega\approx kv_F$ and since Dirac mode decays slowly: $E_D(t)\sim 1/\sqrt{t}$, one might confuse it with a lossy plasmon. 

Let us show that a similar behavior is observed from the perturbation localized in space and periodic in time:
\begin{equation}
E^{ext}(x,t)=-\varPhi\delta(x)\cos\omega t.
\end{equation}
This also corresponds to a more typical experimental setup used to excite plasmons via scanning near field optical microscope (SNOM) \cite{Chen12,Fei12}.
Note that $\varPhi=-\int E^{ext}(x,0)dx$ corresponds to the amplitude of the scalar potential (now in the gauge $A=0$, $\varPhi\neq 0$). From Eq. (\ref{screenE}) we can find the total field:
\begin{equation}
E(x,t)=-\frac{\varPhi}{2} e^{-i(\omega+i\eta)t}\int_{-\infty}^\infty\frac{dk}{2\pi}\frac{e^{ikx}}{\varepsilon(k,\omega+i\eta)}+c.c.,
\label{intK}
\end{equation}
where $c.c.$ stands for the complex conjugate. We have again introduced positive imaginary part to the frequency and we have used a general relation: $\varepsilon(-\omega^*)=\varepsilon^*(\omega)$ \cite{LLPK}. Also for the sake of convenience, instead of $k_x$ here we write $k$ which can now be positive and negative. One should just keep in mind that in this quasi 1D case $\varepsilon(k)$ depends only on the magnitude of wavenumber $|k|$.

Same as before, integrand in Eq. (\ref{intK}) has singularities at the plasmon poles $k=\pm k_P$ where ${\varepsilon(k,\omega+i\eta)=0}$, and at points $k=\pm(\omega+i\eta)/v_F$ which correspond to singularities of $\varepsilon(k,\omega+i\eta)$. If $\omega>0$ then these points are pushed slightly into a lower half of the complex $k$ plane on negative real semiaxis ($\Re k<0$) and oppositely on positive semiaxis ($\Re k>0$). To evaluate the integral (\ref{intK}) for $x>0$ we close the integration contour in the upper half of the complex $k$ plane (see Fig. \ref{fig}(a)). Result is simply given by the singularities of the function $1/\varepsilon(k,\omega)$, which are given by the zeros of $\varepsilon(k,\omega)$ (i.e. the plasmon poles) and the singularities of the $\varepsilon(k,\omega)$ which we specify by putting a branch cut along the lines $\langle-\infty,-\omega/v_F\rangle$ and $\langle\omega/v_F,\infty\rangle$. 
Considering the plasmon poles, only the pole on the positive real semiaxis ($k=+k_P$) contributes to the integral (\ref{intK}), which is easily evaluated by the residuum theorem:
\begin{equation}
E_P(x,t)=\varPhi\ \frac{\sin(k_Px-\omega t)}{\partial\varepsilon/\partial k |_{k_P}}.
\end{equation}

It is clear that only the plasmon pole with positive wavenumber $k=+k_P$ contributes for $\omega>0$, since at $x>0$ we can only have a plasmon propagating to the right of the source. 
Again we see that the plasmon emission is suppressed for $\chi_{k_P}\ll 1$ since from Eq. (\ref{difu}):
\begin{equation}
\left.\frac{\partial\varepsilon}{\partial k}\right\vert_{k_P}\approx-\frac{1}{\chi_{\omega/v_F}^2 \omega/v_F},
\end{equation}
where we took into account that for acoustic plasmon: $k_P\approx\omega/v_F$. Plasmon field in this limit is:
\begin{equation}
E_P(x,t)\approx-\chi_{\frac{\omega}{v_F}}^2 \frac{\omega}{v_F}\varPhi\sin\left[\frac{\omega}{v_F}(x-v_Ft)\right].
\end{equation}

The remaining part of the integral (\ref{intK}) is given by integral along the cut line $\langle\omega/v_F,\infty\rangle$, which we again refer to as the Dirac mode. In this case we were able to obtain analytical solution only for $\varepsilon_k=\bar\varepsilon$ i.e. when there is no metal plate, so we present this case only.
Strictly speaking there is then also a singularity at $k=0$ where $\varepsilon(k,\omega)\rightarrow\infty$, but it doesn't contribute to integral (\ref{intK}) since $1/\varepsilon=0$.
From Eq. (\ref{noscreen}) we get the Dirac mode:
\begin{equation}
E_D(x,t)\approx\varPhi e^{-i\omega t}\int_{\frac{\omega}{v_F}}^\infty\frac{dk}{2\pi}\frac{\frac{\omega}{kv_F}\chi_{\frac{\omega}{v_F}}i\omega e^{ikx}}{\sqrt{k^2v_F^2-\omega^2}}+c.c..
\end{equation}
Then using the relation: $ie^{ikx}/k=-\int_0^x e^{ikx}dx+i/k$, and integral representation of the Hankel function of the first kind $H^{(1)}_\nu$ we obtain \cite{AS}:
\begin{equation}
\begin{split}
&E_D(x,t)\approx\chi_{\frac{\omega}{v_F}}\frac{\omega}{v_F}\frac{i}{4}\varPhi e^{-i\omega t}
\times\\
&\times\left(-\int_0^{\frac{x\omega}{v_F}}H_0^{(1)}(\xi)d\xi+1\right) +c.c.,
\end{split}
\label{diracx}
\end{equation}
which can be further expressed via the Struve functions $H_\nu$ using the relation \cite{AS}:
\begin{equation}
\begin{split}
&\int_0^xH_0^{(1)}(\xi)d\xi=xH_0^{(1)}(x)+\frac{\pi}{2}x \times \\
&\times\left( H_0(x)H_1^{(1)}(x)-H_1(x)H_0^{(1)}(x) \right).
\end{split}
\end{equation}

To get the field at $x<0$ we close the integration contour in the lower half of the complex $k$ plane, and one can easily show that $E_D(x<0,t)=E_D(-x,t)$.

Again we see that plasmon emission is suppressed by a factor $\chi_{\omega/v_F}\ll 1$ compared to the Dirac mode. Note the oscillating, slowly decaying behavior of the Dirac mode (plotted in Fig. \ref{fig}(b)). Experimentalists are again warned not to confuse it with a lossy plasmon. 

\section{Echo emission}

Let us first explore temporal echoes and look at the system response to two pulses:
\begin{equation}
E^{ext}(x,t)=-A_1\delta(t)\cos k_1x-A_2\delta(t-\tau)\cos k_2x.
\end{equation}

Again we focus at the regime of acoustic plasmons $\chi_k\ll 1$ when the plasmon emission is weak. We can then neglect the action of the screening field on the system and simply write the total field $E=E^{ext}$ (mathematically this corresponds to Eq. (\ref{noscreen})). It is then particularly simple to solve the Vlasov Eq. (\ref{vlasov}) in the perturbative approach. To do so we look at the general problem:
\begin{equation}
0=\frac{\partial f}{\partial t}+v_x\frac{\partial f}{\partial x}-\delta(t-\tau)a(x).
\label{pulset}
\end{equation}
Using the Fourier transformations we obtain for $t>\tau$:
\begin{equation}
f(x,t)=a(x-v_x(t-\tau)),
\label{wavet}
\end{equation}
which could have been anticipated as particles freely evolve with velocities $v_x$ after the pulse has ceased.
Specifically the linearized Vlasov equation response to the first pulse is: 
\begin{equation}
0=\frac{\partial\delta f_1}{\partial t}+v_x\frac{\partial\delta f_1}{\partial x}+\delta(t)A_1e\cos k_1x\frac{\partial f_0}{\partial p_x},
\end{equation}
where: $f-f_0=\delta f_1\propto A_1$. We thus obtain:
\begin{equation}
\delta f_1(x,t)=-A_1e\frac{\partial f_0}{\partial p_x}\cos k_1(x-v_xt),
\label{freeD}
\end{equation}
which corresponds to the Dirac mode from Eq. (\ref{diract}) that we calculated by a different approach. Compare also to Eq. (\ref{free1}) from the introduction.

Next we include the second pulse and look at the solution in the form: $f=f_0+\delta f_1+\delta f_2$. The Vlasov equation linearized in the variable $\delta f_2\propto A_2$ is given by:
\begin{equation}
0=\frac{\partial\delta f_2}{\partial t}+v_x\frac{\partial\delta f_2}{\partial x}+\delta(t-\tau)A_2e\cos k_2x\frac{\partial(f_0+\delta f_1)}{\partial p_x}.
\end{equation}

Here the term containing $\partial f_0/\partial p_x$ leads to a decaying Dirac mode just like the response to the first pulse given by Eq. (\ref{freeD}), which is irrelevant since it doesn't contribute to echo. Moreover at large times $t\cdot k_1v_F\gg 1$, the derivative in $\partial\delta f_1/\partial p_x$ affects only the cosine in Eq. (\ref{freeD}) since this gives a large factor $\propto t$. We are thus left with the following equation:
\begin{equation}
\begin{split}
0=\frac{\partial\delta f_2}{\partial t}+v_x\frac{\partial\delta f_2}{\partial x}-\delta(t-\tau)A_1A_2e^2\times\\
\times\cos k_2x\sin k_1(x-v_x\tau)k_1\tau\frac{\partial v_x}{\partial p_x}\frac{\partial f_0}{\partial p_x}.
\end{split}
\end{equation}

Then using equations (\ref{pulset}) and (\ref{wavet}) we obtain:
\begin{equation}
\begin{split}
&\delta f_2(x,t)=
A_1A_2e^2k_1\tau\frac{\partial v_x}{\partial p_x}\frac{\partial f_0}{\partial p_x}\times\\
&\times\cos k_2(x-v_x(t-\tau))\sin k_1(x-v_xt).
\end{split}
\end{equation}
Compare it also to Eq. (\ref{free2}) from the  introduction. Just like there we again write the product $\cos\cdot\sin$ as the sum of two sines, and select only the one which contributes to the echo:
\begin{equation}
\begin{split}
\delta f_2^{echo}(x,t)=
A_1A_2e^2k_1\tau v_F\frac{1}{2p}\frac{df_0}{dp}\sin^2\theta\cos\theta\times\\
\times\sin (-k_3x+(k_3t-k_2\tau)v_F\cos\theta),
\end{split}
\end{equation}
where $k_3=k_2-k_1$. We can then calculate the induced charge density $\rho^{ind}=-e\int d{\bf p}\delta f_2^{echo}$ and using Eq. (\ref{indE}) the induced (echo) field that can be measured:
\begin{equation}
\begin{split}
&E^{echo}(x,t)=
A_1A_2\frac{e^3k_1\tau v_F}{4\varepsilon_0\varepsilon_{k_3}}f_0(0)\times\\ &\times\int_0^{2\pi}d\theta\sin^2\theta\cos\theta
\cos(-k_3x+(k_3t-k_2\tau)v_F\cos\theta).
\end{split}
\end{equation}

Finally using the integral representation of Bessel functions and the relation \cite{AS}: $J_1(\xi)+J_3(\xi)=4J_2(\xi)/\xi$, we obtain explicit expression for the echo electric field:
\begin{equation}
\begin{split}
E^{echo}(x,t)=
A_1A_2\frac{\pi e^3k_1\tau v_F}{2\varepsilon_0\varepsilon_{k_3}}f_0(0)
\times\\
\times\sin k_3x\ \frac{J_2(v_F(k_3t-k_2\tau))}{v_F(k_3t-k_2\tau)}.
\label{echotE}
\end{split}
\end{equation}

One can clearly see a strong echo response near the time $\tau_{echo}$ given by Eq. (\ref{echot}) and a decay in $E^{echo}$ as we move away from that point in time. Echo field amplitude is of the order:
\begin{equation}
E^{echo}\sim A^2\chi_kk^2v_F^2\tau e/p_F
\end{equation}
where we assumed that $A_{1,2}\sim A$, $k_{1,2}\sim k$ and $T=0$ K. First of all we can see that echo will become comparable to the plasmon field (\ref{plasmont}) if:
\begin{equation}
A\sim\frac{p_F}{e}\frac{\chi_k}{kv_F\tau},
\label{Ap}
\end{equation}
which is quite low since $\chi_k\ll 1$ and $kv_F\tau\gg1$. Somewhat larger field is required to have echo comparable to the Dirac mode amplitude (\ref{diract}):
\begin{equation}
A\sim\frac{p_F}{e}\frac{1}{kv_F\tau},
\label{AD}
\end{equation}

One might worry here about other nonlinear effects that we disregarded in our analysis. For example it was shown that acoustic plasmons have extremely large nonlinear response \cite{Jablan20}, so one might wonder if we can really disregard plasmon/screening field in the nonlinear case (strictly speaking we have demonstrated that this is true only in the linear response). Specifically it was shown that linear response breaks down for the plasmon vector potential amplitude $A_P\sim(1-kv_F/\omega)p_F/e$ \cite{Jablan20}. One can get a rough understanding of this result by looking into two opposite regimes. In the homogenous case $k=0$ equation of motion is easily solved by introducing the generalized momentum ${\bf p}\rightarrow{\bf p}-e{\bf A}_P$. Then for $eA_P\sim p_F$ our field will probe the Dirac point singularity resulting in the nonlinear behavior. On the other hand in the resonant case $\omega\approx kv_F$ linear response function $\varepsilon(q,\omega)$ diverges indicating the breakdown of the linear response regime \cite{Jablan20}. Particularly we can obtain plasmon potential amplitude from Eq. (\ref{plasmont}): $A_P=2\chi_k^2A$, since $E_P(x,t)=-\partial A_P(x,t)/\partial t$. Finally as $1-kv_F/\omega\sim\chi_k^2$ from Eq. (\ref{difu}), we find the amplitude needed for the plasmon to enter nonlinear response: $A\sim p_F/e$, which is larger than the fields given by expressions (\ref{Ap}) and (\ref{AD}). This further justifies our assumption to neglect the nonlinear screening field. 

One should also note that the echo response is an effect of the order $\sim E_1E_2$, but we disregarded terms of the order $\sim E_{1,2}^2$ since they influence echoes only as a higher order effect. 
However these terms will lead to Dirac modes at the second harmonics $2k_{1,2}$ which should be simple to observe in experiments.

Since it is probably not easy to obtain a temporal pulse perfectly harmonic in space it might be more convenient to pattern the graphene surface with a periodically alternating dielectric (1D photonic crystal) and couple the normal incidence light into spatial harmonics.
If $D$ is the period of the photonic crystal then light will excite all the harmonics with wavenumber $k_\nu=2\nu\pi/D$. Our former echo analysis can then be easily generalized by using a periodic function $a(x)$ in Eq. (\ref{pulset}) instead of a simple $\cos kx$. Moreover one could use Eq. (\ref{echotE}) to get a lowest order echo estimate by writing $k_2=2k_1=2k_3=4\pi/D$, however one first has to solve the exact scattering problem and relate the amplitudes $A_{1,2}$ and $E^{echo}$  to the input and output light intensities. 

It probably simpler to use a monochromatic light (harmonic in time) which can be localized in space by hitting a SNOM tip for example. 
Let us then explore spatial echo and look at the system response to a field:
\begin{equation}
E^{ext}(x,t)=-\varPhi_1\delta(x)\cos\omega_1t-\varPhi_2\delta(x-l)\cos\omega_2t.
\label{pertx}
\end{equation}

Like before we study the general problem of a type:
\begin{equation}
0=\frac{\partial f}{\partial t}+v_x\frac{\partial f}{\partial x}-\delta(x-l)\phi(t).
\label{pulsex}
\end{equation}
This case being more tricky we give all the steps of the analysis. We start by finding the Fourier components:
\begin{equation}
f_{k\omega}=-i\frac{\phi_\omega}{v_x}\frac{e^{-ikl}}{k-\frac{\omega+i\eta}{v_x}}.
\end{equation}
Performing first the Fourier transform in space:
\begin{equation}
f_\omega(x)=\int\frac{dk}{2\pi}e^{ikx}f_{k\omega}=-\frac{i\phi_\omega}{2\pi v_x}\int dk\frac{e^{ik(x-l)}}{k-\frac{\omega+i\eta}{v_x}}.
\end{equation}
For $x>l$ we close the integration contour in the upper part of the complex $k$ plane but note that we pick up the singularity at $k=\omega/v_x$ only if $v_x>0$. Using the residuum theorem we thus obtain:
\begin{equation}
f_\omega(x)=\frac{\Theta(v_x)}{v_x}\phi_\omega e^{i(x-l)\omega/v_x},
\end{equation}
where $\Theta(x)$ is a unit step function. Finally we perform the Fourier transform in time to obtain:
\begin{equation}
f(x,t)=\frac{\Theta(v_x)}{v_x}\phi\left(t-\frac{x-l}{v_x}\right).
\end{equation}
Note a divergence in the response at $0=v_x=v\cos\theta$.

The rest of the calculation perfectly parallels the case of temporal echoes so we only give the final result valid for a large distance $l\omega_1/v_x\gg1$:
\begin{equation}
\begin{split}
&E^{echo}(x,t)=-\varPhi_1\varPhi_2\frac{e^3\omega_1 l }{4\varepsilon_0\bar\varepsilon}\times\\ 
&\times\int d{\bf p}\frac{\Theta(v_x)}{v_x^4}\frac{\partial v_x}{\partial p_x}\frac{\partial f_0}{\partial p_x}
\cos\left(-\omega_3t+\frac{x\omega_3-l\omega_2}{v_x}\right),
\end{split}
\label{echoxEintgen}
\end{equation}
where $\omega_3=\omega_2-\omega_1$ and we assumed that there is no metal plate i.e. $\varepsilon_k=\bar\varepsilon$. Particularly in graphene we get:
\begin{equation}
\begin{split}
&E^{echo}(x,t)=\varPhi_1\varPhi_2\frac{e^3\omega_1l}{2\varepsilon_0\bar\varepsilon v_F^3}f_0(0)
\times\\
&\times\int_{0}^{\frac{\pi}{2}}d\theta\frac{\sin^2\theta}{\cos^3\theta}
\cos\left(-\omega_3t+\frac{x\omega_3-l\omega_2}{v_F\cos\theta}\right).
\label{echoxEint}
\end{split}
\end{equation}
Then by making a substitution: $\cosh\zeta=1/\cos\theta$, using the integral representation of the Hankel function and the relation \cite{AS}: $\ddot{H}^{(1)}_0(\xi)+H^{(1)}_0(\xi)=H^{(1)}_1(\xi)/\xi$, we obtain:
\begin{equation}
\begin{split}
E^{echo}(x,t)=-\varPhi_1\varPhi_2\frac{\pi e^3\omega_1l}{8\varepsilon_0\bar\varepsilon v_F^3}f_0(0)e^{-i\omega_3t}\times\\
\times\frac{iH^{(1)}_1\left(\left|\frac{x\omega_3-l\omega_2}{v_F}\right|\right)}{\left|\frac{x\omega_3-l\omega_2}{v_F}\right|}+c.c.,
\end{split}
\end{equation}
which is plotted in Fig. \ref{fig}(b) at $t=0$ s. We can clearly see a large echo response at the position:
\begin{equation}
x=l_{echo}=l\cdot\frac{\omega_2}{\omega_2-\omega_1},
\end{equation}
and a decay as we move away from that point. For $x<l_{echo}$ echo field is given by the same formula only $iH^{(1)}_1$ should be replaced by $-iH^{(1)*}_1$.

To get a sense of the strength of nonlinear echo response we assume $T=0$ K, and introduce the system scale parameter:
\begin{equation}
K=k\chi_k=\frac{4\pi e^2\mathscr{E}_F}{\varepsilon_0\bar\varepsilon h^2v_F^2}.
\end{equation}
The scale of the echo field can then be written as:
\begin{equation}
E^{echo}\sim \frac{\varPhi^2Ke}{\mathscr{E}_F}\frac{\omega_1l}{v_F}.
\end{equation}
Let us compare this field to the amplitude of the Dirac mode from Eq. (\ref{diracx}): $E_D\sim \varPhi K$. We can make these two effects comparable: $E^{echo}\sim E_D$, for $E_D\sim (K\mathscr{E}_F/e)/(\omega_1l/v_F)$. To make sense of this field we compare it to the intrinsic electric field that naturally comes about in the study of nonlinear effects in graphene \cite{Jablan15}:
\begin{equation}
E_{e-e}=\frac{e}{4\pi\varepsilon_0\bar\varepsilon r_e^2}.
\end{equation}
This is just the field between two electrons at an average distance $r_e$ given by: $r_e^2\pi=1/n$. Since the electron density is: $n=\int d{\bf p}f_0=4\pi p_F^2/h^2$, we can write: $E_{e-e}\sim K\mathscr{E}_F/e$. Finally the required strength of the Dirac field:
\begin{equation}
E_D\sim \frac{E_{e-e}}{\frac{\omega_1l}{v_F}},
\end{equation}
so we can lower the nonlinear threshold by using a large distance $l$ between our sources.

Note that spatial echo diverges at $x=l_{echo}$ since ${H^{(1)}_1(\xi)\sim1/\xi}$ for small $\xi\ll 1$, which is not the case for temporal echoes where $J_2(\xi)\sim \xi^2$ \cite{AS}. This clearly points that our perturbative analysis of spatial echoes breaks apart near the point $l_{echo}$. 
Note also that this divergence is not removed by the screening field, which is easily seen by dividing the integrand in Eq. (\ref{echoxEint}) by a dielectric function from Eq. (\ref{difu}):
\begin{equation}
\begin{split}
\varepsilon(\theta)&=1+\chi_{\frac{\omega_3}{v_F\cos\theta}}\left(1-\frac{1}{\sqrt{1-1/\cos^2\theta}}\right)\\
&=1+\cos\theta\chi_{\frac{\omega_3}{v_F}}\left(1+i\frac{\cos\theta}{\sin\theta}\right).
\end{split}
\end{equation}
This only reduces the regime of small angles $\theta\approx0$ which is in any case small due to a factor $\sin^2\theta$ (again justifying our assumption to neglect the screening field), while the problematic regime is around $\theta\approx\pi/2$ where $1/\cos^3\theta\rightarrow\infty$ (and screening is negligible $\varepsilon\approx1$). Similar arguments are valid for the case of parabolic electron dispersion in 2D, while in 3D spatial echoes don't show this divergent response \cite{ONeil68}. This is a consequence of a different nature of the Coulomb field in 3D since: $\nabla {\bf E}=\rho^{3D}/\varepsilon_0$, i.e.
\begin{equation}
E^{3D}_k=-i\frac{\rho^{3D}_k}{k\varepsilon_0}.
\label{E3D}
\end{equation}
Compare also to Eq. (\ref{indE}) in 2D. Again the issue is with large angles $\theta\approx\pi/2$ i.e. small velocity regime $v_x$ i.e. large wavenumbers $k=\omega_3/v_x$ in Eq. (\ref{echoxEintgen}).
So it is really this slow (i.e. fast) decay of the Coulomb field at large wavenumbers that give the divergent (finite) echo field in 2D (3D).
Divergence of a 2D case points to a strong echo response, but one can't extrapolate the actual echo amplitude at the position $l_{echo}$ with the perturbative approach presented here. In fact echo amplitude will be strongly influenced by electron collisions beyond mean field, since this divergence can be traced down to the small velocity $v_x$ regime. Namely for these slow electrons to influence the echo they have to be able to travel the distance of the order $\sim l$ before they get scattered i.e. $v_x\gtrsim l/\tau_{col}$, which introduces a low velocity cutoff and removes the divergence issue. Here $\tau_{col}$ is the large angle collision time required to kick the particle out of it's trajectory (see also discussion below).

\section{Discussion}

While in this article we focus on plasma echoes in graphene since this is the most interesting problem from theoretical perspective (linear dispersion not been treated before) we urge experimentalist to study the general case of these new 2D crystals \cite{Novoselov05}. In fact the parabolic electron dispersion might be more suitable than the linear dispersion to study temporal echoes. While both work fine in the small momentum limit, only parabolic dispersion will work at large momenta ($\hbar k\sim p_F$). This can be seen by calculating the quantum mechanical response of the system to two pulses at wavevectors ${\bf k}_1$ and ${\bf k}_2$ separated by a time $\tau$. After straightforward calculations we obtain expression for the induced field (at the wavevector ${\bf k}_3={\bf k}_2-{\bf k}_1$) in the second order response of the form:
\begin{equation}
E^{(2)}_{{\bf k}_3}(t) \sim \int d{\bf k}  \\\ 
e^{i(\mathscr{E}_{\bf k}-\mathscr{E}_{{\bf k}+{\bf k}_3})t/\hbar} \\\
e^{i(\mathscr{E}_{{\bf k}+{\bf k}_2}-\mathscr{E}_{\bf k})\tau/\hbar}.
\end{equation}

Specifically at large times $t$ and $\tau$, any $\bf k$-dependence in the exponent will lead to large cancelations of the integral $\int d\bf k$. However it might happen at a certain time $t=\tau_{echo}$ that the exponent looses the $\bf k$-dependence returning the finite $E^{(2)}_{{\bf k}_3}$ response. To find the exact condition for this echo to arise we can compare the successive terms in the Taylor expansion of the energy differences: $\mathscr{E}_{\bf k}-\mathscr{E}_{{\bf k}+{\bf k}_\nu}$. In the small momentum limit it is sufficient to take the lowest order $\sim k_\nu$ to find the echo time: $\tau_{echo}=\tau k_2/k_3$ (assuming ${\bf k}_\nu =k_\nu{\bf \hat{x}}$). Thus in the small momentum (long wavelength) limit echo appears regardless of the exact electron dispersion. However for larger momenta we need to look in the next order $\sim k_\nu^2$.
If we tried to vary first and second order independently we would conclude that the echo also has to satisfy: $\tau_{echo}=\tau k^2_2/k^2_3$, which is obviously impossible. One way to go beyond small momentum is to have $\partial^2\mathscr{E}/\partial k^2_x$ as a $\bf k$-independent constant. In other words we need a parabolic energy dispersion $\mathscr{E}_{\bf k}=\hbar^2k_x^2/2m+\mathscr{E}'_{k_y}$ (with arbitrary dispersion in the $y$-direction). It is easy to check that this specific dispersion indeed gives echo response for any $k_1$ and $k_2$ ($k_3=k_2-k_1$). 
Another way would be to have $\partial^2\mathscr{E}/\partial k_x^2\propto\partial \mathscr{E}/\partial k_x$ which is satisfed by an exponential dispersion: $\mathscr{E}_{\bf k}=\mathscr{E}_0e^{lk_x}+\mathscr{E}'_{k_y}$. It is easy to see that we again get echo for any $k_{1,2}$ but this dispersion doesn't look very physical. 

However what will really matter in the end is how sensitive are these echoes to collisions (think of the horse collisions from the introduction). In fact it was shown that electron-electron (e-e) collisions have large influence on the echoes due to rapid oscillations of the distribution function, and it is useful to repeat the argument in the 3D case \cite{ONeil68,LLPK}. 
Since e-e interaction is long range, small angle collisions are very important and sufficient to smooth out fine scale momentum oscillations of the echo distribution. Furthermore as there are many more small than large angle Coulomb collisions, echo could vanish even if the echo time is much smaller than the large angle collision time: $\tau_{echo}\ll\tau_{col}$  \cite{ONeil68}. From a different perspective, small angle collisions can be described by small change of the momenta and thus lead to a diffusion behavior described by a gradient of the particle flux in the momentum space: $df/dt=-\nabla_{\bf p}{\bf s}$. Fast oscillations of echo distribution in momentum space then lead to large increase in e-e collision \cite{LLPK}. 

On the other hand Coulomb interaction $U(r)\propto1/r$ in 2D is less singular in the small momentum transfer as the Fourier transforms are: $U^{2D}_k\propto 1/k$ (corresponding to Eq. (\ref{induced})) and $U^{3D}_k\propto 1/k^2$ (corresponding to Eq. (\ref{E3D})). One might hope that e-e collisions are then less effective in reducing 2D echoes since the scattering cross section in the Born approximation is: $d\sigma\propto|U(k)|^2$ \cite{LLQM}, but careful analysis of the problem is needed. 
For 2D spatial echoes the problem is even more intricate as we saw that echo field diverges (due to a slow decay of the Coulomb field at large wavenumbers) and the large angle collisions strongly influence the echo amplitude. 
We also emphasize that in our case the large substrate screening (required to reach the acoustic plasmon dispersion) has an extra benefit of reducing these collisions thus increasing the echo response. 

Before closing we would like to note that the interaction of plasmons and echoes is an intricate mathematical problem. For example in the parabolic case plasmon experience Landau damping and while Landau performed his calculations in the linearized regime \cite{LLPK}, full nonlinear Landau damping was only recently solved \cite{Villani14}. Major problem was to show that echoes do not accumulate constructively to a massive response. Echoes there played a role similar to the dangerous resonances in the Kolmogorov-Arnold-Moser theory. It would be very interesting to see if the same approach could handle echoes in the case of linear dispersion in graphene which adds additional singularities (in electron dispersion at the Dirac point, and in dielectric function at $\omega=kv_F$). However it seems that mathematical rigor is still not readily obtainable in theoretical physics, as noted by Landau so many years ago \cite{LLSP}. Most likely this subtle long time perturbative behavior discussed in reference \cite{Villani14} will be overwhelmed by electron collisions.

In conclusion we have obtained analytic expressions for temporal and spatial plasma echoes in graphene when (acoustic) plasmon emission is suppressed. 
We found extremely strong spatial echo response and discussed how electron collisions reduce the echo. We also discussed differences between various electron dispersions, and differences between semiclassical and quantum model of echoes. 
Optimistically looking one could hope to use these nonlinear echoes for all optical signal processing like optical switch or optical memory (since the result is delayed). Most likely echoes could be used as a sensitive probe of electron interactions in 2D crystals.



This work was supported by QuantiXLie Centre of Excellence, a project cofinanced by the Croatian Government and European Union through the European Regional Development Fund - the Competitiveness and Cohesion Operational Programme (Grant KK.01.1.1.01.0004).


\end{document}